\documentclass[floatfix,5p]{elsarticle}

\usepackage{bm}
\usepackage{braket}
\usepackage{amssymb}
\usepackage{graphicx}
\usepackage{dcolumn}
\usepackage{amsmath}
\usepackage{amsfonts}
\usepackage{xcolor}
\usepackage{hyperref}

\biboptions{numbers,sort&compress}
\usepackage{mathtools}
\mathtoolsset{showonlyrefs}

\hypersetup{
    colorlinks,
    linkcolor={red},
    citecolor={blue},
    urlcolor={green}
}

\begin{document}

\title{Investigating the weak charge of $^{48}$Ca using a dispersive optical model}
\address[1]{Department of Physics, Washington University in St. Louis, MO 63130 USA}
\address[2]{Lawrence Livermore National Laboratory, P.O. Box 808, L-414, Livermore, CA 94551, USA}
\author[1]{N. L. Calleya\corref{cor1}}
\ead{calleya@wustl.edu}
\cortext[cor1]{Corresponding author}
\author[2]{M. C. Atkinson} 
\author[1]{W. H. Dickhoff} 
\date{\today}

\begin{abstract}
   A new nonlocal dispersive-optical-model analysis has been carried out for neutrons and protons in $^{48}$Ca that reproduces the weak-form-factor measurement of CREX. In addition to elastic-scattering angular distributions, total and reaction cross sections, single-particle energies, the neutron and proton numbers, and the charge distribution, the CREX-measured weak form factor has been fit to extract the neutron and proton self-energies both above and below the Fermi energy.  The resulting single-particle propagators yield a weak form factor of $F_W = 0.125 \pm 0.05$ and a neutron skin of $R_\mathrm{skin} = 0.152 \pm 0.05$ fm, in good agreement with CREX. 
The rearrangement of the neutron distribution to accommodate such a thin neutron skin results in the high-momentum content of the neutrons exceeding that of the protons, in contrast to what is expected from high-energy two-nucleon knockout measurements by the CLAS collaboration and \textit{ab initio} asymmetric matter calculations. The present analysis also emphasizes the importance of neutron experimental data in constraining weak charge observables necessary for a precise description of neutron densities. Notably, the neutron reaction cross section and further parity-violating experiments weak form factor measurements are essential to generate a unique way to determine the $^{48}$Ca neutron distribution in this framework.
\end{abstract}

\maketitle


In contrast to proton distributions and its related charge density, which has been experimentally probed for many nuclei across the periodic table, the neutron counterpart remains elusive. In particular, for nuclei with more neutrons than protons such as $^{48}$Ca, the determination of how these nucleons are distributed over the nuclear volume is of significant importance not only for its value as a nuclear-structure observable but also for its close relation to the neutron skin thickness $R_\mathrm{skin}$ and consequently what this quantity represents. 

Defined as the difference between neutron and proton root-mean-squared (RMS) radii, \textit{i.e.}, $R_\mathrm{skin} = R_{n} – R_{p}$, the neutron skin has a connection to the nuclear symmetry energy: its value is determined by the relative strengths of the symmetry energy between the central near-saturation and peripheral less-dense regions. In other words, $R_\mathrm{skin}$ is a measure of the density dependence of the symmetry energy around saturation~\cite{Typel01,Furnstahl02,Steiner05,RocaMaza11}. The skin is also directly correlated to the slope of the symmetry energy $L$, since a thicker skin favors a larger $L$ and a thinner skin a smaller one. 
The precise determination of $R_\mathrm{skin}$ is pertinent to further understand many nuclear properties including masses, radii, and the location of the drip lines in the chart of nuclides. Its importance extends to astrophysics for understanding supernovae and neutron stars~\cite{Horowitz01,Steiner10} due to the aforementioned symmetry energy role in the equation of state (EOS) for nuclear matter, and to heavy-ion reactions~\cite{li08}.

Given the broad implications in a wide variety of physics research areas, the neutron skin has been the subject of many studies both experimental and theoretical to determine its thickness~\cite{Tsang12,Mammei:2024}.  While $R_p$ is extracted quite accurately from the charge form factor, $F_\mathrm{ch}$, derived from elastic electron scattering cross sections~\cite{Angeli:2013} and laser spectroscopy~\cite{Garcia:2016}, most experimental determinations of $R_n$ are model dependent~\cite{Tsang12} and rely on strong interacting probes. With the use of parity-violating electron scattering~\cite{Horowitz98}, it is possible to obtain the weak distribution with the same degree of model independence since the weak charge distribution. Since the weak charge distribution is predominantly determined by the neutrons, this is an accurate method for determining the neutron skin. 

The first parity-violating experiment performed by the PREX collaboration yielded a thick neutron skin of $^{208}$Pb with a rather large uncertainty~\cite{PREX12}. A second experiment, dubbed PREX-2, was later performed resulting in a $^{208}$Pb skin of $R_\mathrm{skin}^{208} = 0.283 \pm 0.071$ fm~\cite{prex2:2021}. The following year, the CREX experiment extracted a much smaller skin in $^{48}$Ca of $R_\mathrm{skin}^{48} = 0.121 \pm 0.026(\mathrm{exp})\pm 0.024(\mathrm{model})$ fm~\cite{crex:2022}.
A thick-thin skin scenario between the two asymmetric nuclei creates friction regarding the nuclear EOS and slope of the symmetry energy $L$, and also in exotic astrophysical systems such as neutron stars~\cite{Piekarewicz2024}. More specifically, mass-radius curves predicted from the two different $R_\mathrm{skin}$-derived EOS are incompatible with each other and even with observations. 

Currently there are no adequate models that can predict both PREX-2 and CREX results for neutron skins simultaneously. 
Mean-field approaches predict a strong positive correlation between the neutron skins of $^{208}$Pb and $^{48}$Ca although it has been argued that the large error bars for PREX-2 may not provide a stringent constraint on the isovector part of energy density functionals~\cite{Reinhard:2022}. \textit{Ab initio} approaches also exist for both nuclei. In Ref.~\cite{Hagen:2016} a neutron skin for $^{48}$Ca was predicted that is consistent with CREX while the result of Refs.~\cite{Hu:2022,Hu:2024} exhibits mild tension with PREX-2. 

A unique approach to determining neutron skins in $^{48}$Ca and $^{208}$Pb is provided by the dispersive optical model (DOM) which, unlike mean-field or \textit{ab initio} methods applied to these nuclei, describes scattering observables in addition to bound nucleon properties by making use of a dispersion relation that couples both energy domains above and below the Fermi energy. 
By leveraging Green's function theory, the DOM establishes a nucleon self-energy as a phenomenological optical potential constrained by both bound-state and scattering measurements~\cite{Mahzoon:2014,Atkinson20,Atkinson:2018}.

An earlier DOM analysis of $^{208}$Pb predicted a neutron skin of $R_\mathrm{skin}^\mathrm{DOM}=0.25\pm0.05$ fm which is within 1$\sigma$ of the PREX-2 measurement published the following year~\cite{Atkinson:2020}. The DOM analysis of $^{48}$Ca in 2017 resulted in $R_\mathrm{skin}^\mathrm{DOM}=0.249\pm0.023$ fm which, while consistent with the correlations suggested by previous systematic studies, is over 2$\sigma$ away from the CREX measurement published five years later~\cite{Mahzoon:2017}. These skins are also in agreement with those predicted by a separate DOM fit using a slightly different parametrization of the optical potential together with a Markov Chain Monte Carlo approach~\cite{Pruitt:2020,Pruitt:2020C}. 

We aim to confront the CREX-PREX puzzle by constraining the DOM self-energy to reproduce the CREX-measured weak form factor at momentum transform $q = 0.8733$ $\mathrm{fm}^{-1}$ of $F_W = 0.1304\pm0.0052(\mathrm{stat})\pm0.0020(\mathrm{syst})$. 
Our new DOM fit of $^{48}$Ca results in a much thinner skin of $R_\mathrm{skin}^\mathrm{DOM} = 0.152 \pm 0.05$ fm while maintaining an acceptable reproduction of the previously-fit measurements. While all observables are in good agreement with experimental data, we find that the high-momentum content, defined here as the percentage of particles with momentum greater than 270 MeV/c, is affected in a way consistent with a more confined neutron distribution. This may point to tension with expectations based on high-energy knockout measurements by the CLAS collaboration~\cite{CLAS:2006,Duer:2018} as well as \textit{ab initio} asymmetric matter calculations that properly treat the effect of the nuclear tensor force~\cite{Rios:2009,Rios:2014}. 






In the many-body Green's function formalism the so-called irreducible self-energy, $\Sigma^*(\bm{r},\bm{r}';E)$, is a complex one-body potential which, in principle, is comprised of an infinite set of Feynman diagrams describing the propagation of an interacting nucleon through a nucleus based on a Hamiltonian containing relevant two- and three-body interactions~\cite{Exposed!}.
This complex one-body potential can be parametrized as an optical potential,
 and the link to the negative-energy domain emerges naturally in the Green's function framework as was realized by Mahaux and Sartor who introduced the DOM as reviewed in Ref.~\cite{Mahaux91}. 
 The analytic structure of the nucleon self-energy allows one 
 to apply a dispersion relation, which relates the real part of the self-energy at a given energy to a dispersion integral of its imaginary part over all energies.
 The energy-independent correlated Hartree-Fock (HF) contribution~\cite{Exposed!} is removed by employing a subtracted dispersion relation with the Fermi energy used as the subtraction point~\cite{Mahaux91}
  \begin{align}
    \mathrm{Re}\ \Sigma^*(\alpha,\beta;E) &= \mathrm{Re}\
    \Sigma^*(\alpha,\beta;\varepsilon_F) \label{eq:dispersion} \\ -
    \mathcal{P}\int_{\varepsilon_F}^{\infty} \!\! \frac{dE'}{\pi}&\mathrm{Im}\
    \Sigma^*(\alpha,\beta;E')\left[\frac{1}{E-E'}-\frac{1}{\varepsilon_F-E'}\right] \nonumber
    \\ + \mathcal{P} \! \int_{-\infty}^{\varepsilon_F} \!\!
    \frac{dE'}{\pi}&\mathrm{Im}\
    \Sigma^*(\alpha,\beta;E')\left[\frac{1}{E-E'}-\frac{1}{\varepsilon_F-E'}\right],
    \nonumber      
 \end{align}
 where $\varepsilon_F = \frac{1}{2}(E^{A+1}_0 - E^{A-1}_0)$ is the average Fermi energy which separates the particle and hole domains~\cite{Exposed!}.
 
 The subtracted form has the further advantage that the emphasis is placed on energies closer to the Fermi energy for which more experimental data are available.
 The real part of the self-energy at the Fermi energy is then still referred to as the HF term, $\Sigma_{\text{HF}}$,  but is sufficiently attractive for binding.
 In practice, the imaginary part is assumed to extend to the Fermi energy on both sides while being very small in its  vicinity.
 Initially, standard functional forms for these terms were introduced by Mahaux and Sartor who also cast the DOM potential in a local form by a standard transformation which turns a nonlocal static HF potential into an energy-dependent local potential~\cite{Perey:1962}.
 Such an analysis was extended in Refs.~\cite{Charity06,Charity:2007} to a sequence of Ca isotopes and in Ref.~\cite{Mueller:2011} to semi-closed-shell nuclei heavier than Ca.
 
 The transformation to the exclusive use of local potentials precludes a proper calculation of nucleon particle number and expectation values of the one-body operators, such as the charge density in the ground state. 
 This obstacle was eliminated in Ref.~\cite{Dickhoff:2010}, but it was shown that the introduction of nonlocality in the imaginary part was still necessary in order to accurately account for particle number and the charge density~\cite{Mahzoon:2014}.
 Theoretical work provided further support for this introduction of a nonlocal representation of the imaginary part of the self-energy~\cite{Waldecker:2011,Dussan:2011}.
 A review has been published in Ref.~\cite{Dickhoff:2017}.

We implement a nonlocal representation of the self-energy following
 Ref.~\cite{Mahzoon:2014} where $\Sigma_{\text{HF}}(\bm{r},\bm{r'})$ and the imaginary part 
 $\mathrm{Im}\ \Sigma(\bm{r},\bm{r'};E)$ are parametrized, and Eq.~\eqref{eq:dispersion} generates the energy dependence of the
 real part. The HF term consists of a volume term, spin-orbit term,  and a wine bottle shape~\cite{Brida11} to simulate a surface contribution. The imaginary self-energy consists of volume, surface, and spin-orbit terms. 
 Nonlocality is represented using the Gaussian form  as proposed in Ref.~\cite{Perey:1962}. This form is particularly useful as it has an analytic expression in a partial-wave basis~\cite{Mahzoon:2015}. More details can be found in~\cite{Atkinson:2020}, a description of the potential terms as well the final parameter set is left to the supplementary material.


To use the DOM self-energy for predictions, the parameters are fit through a weighted $\chi^2$ minimization of available elastic differential cross section data ($\frac{d\sigma}{d\Omega}$), analyzing power data ($A_\theta$),  reaction cross sections ($\sigma_r$), total cross sections ($\sigma_t$), charge density ($\rho_{\text{ch}}$), energy levels ($\varepsilon_{\ell j}$), particle number, separation energies,  the root-mean-square charge radius ($R_\mathrm{ch}$), and the energy of the ground state~\cite{Atkinson:2020b}. Also included in this fit is the weak form factor of $^{48}$Ca, $F_W^{48}$. The scattering calculations are performed using the framework of $R$-matrix theory~\cite{Baye:2010} and the bound-state calculations utilize Green's function formalism.  
 All calculations are done in a Lagrange basis with 30 mesh points, where Legendre polynomials mapped to $r= 0 \rightarrow 12$ fm are used for scattering calculations and Laguerre polynomials are used for bound-state calculations~\cite{Baye_review,Baye:2010}.
 

We employ the Dyson equation to obtain the Green's function, $G_{\ell j}(\alpha,\beta;E)$, from the DOM self-energy, 
 \begin{align}
    G_{\ell j}(\alpha,\beta;E) &= G_{\ell}^{(0)}(\alpha,\beta;E) \nonumber \\ +
    \sum_{\gamma,\delta}&G_{\ell}^{(0)}(\alpha,\gamma;E)\Sigma_{\ell
    j}^*(\gamma,\delta;E)G_{\ell j}(\delta,\beta;E) ,
    \label{eq:dyson}
 \end{align}
 where $G^{(0)}_{\ell}(\alpha,\beta;E)$ corresponds to the free propagator (the Green's function when $\Sigma_{\ell j}^*(\gamma,\delta;E)=0$)
 ~\cite{Exposed!}. The particle number, binding energy, and charge density are all obtained from the so-called hole spectral function which corresponds to the imaginary part of the Green's functions, 
 \begin{equation}
    S^{(p,n)}_{\ell j}(\alpha,\beta;E) = \frac{1}{\pi}\mathrm{Im}\ G^{(p,n)}_{\ell j}(\alpha,\beta;E) .
    \label{eq:spec}
 \end{equation}
The single-particle density distribution can be calculated from the hole spectral function in the following way, 
 \begin{equation}
    \label{eq:charge}
    \rho^{(p,n)}(r) = \frac{1}{4\pi} \sum_{\ell j} (2j+1) \int_{-\infty}^{\varepsilon_F}dE\ S^{(p,n)}_{\ell j}(r,r;E),
 \end{equation}
 where we are now explicitly in coordinate space. The RMS radii of the proton and neutron distributions of Eq.~\eqref{eq:charge} are used to calculate $R_\mathrm{skin}$ as well as the nuclear charge radius, 
 \begin{equation}
    \label{eq:rch}
    R_\mathrm{ch}^2 = R_p^2 + \braket{r_p^2} + \frac{N}{Z}\braket{r_n^2} + \braket{r_\mathrm{DF}^2} + \braket{r_{SO}^2}, 
 \end{equation}
 where $\braket{r_{SO}^2}$ is the spin-orbit contribution calculated according to Ref.~\cite{Atkinson20}, $\braket{r_p^2} = 0.709$ fm$^2$ is the charge radius squared of the proton~\cite{Pohl:2010}, $\braket{r_n^2} = -0.106$ fm$^2$ is the charge radius squared of the neutron~\cite{Filin:2020}, and $\braket{r_\mathrm{DF}^2}$ is the so-called Darwin-Foldy term which is a relativistic correction.
 To obtain the charge density, $\rho_\mathrm{ch}(r)$, we fold the single-particle densities with neutron and proton charge distributions
in addition to calculating their spin-orbit contributions as detailed in Refs.~\cite{Atkinson20,Horowitz:2012}.
To ensure that the proton charge density is consistent with $R_\mathrm{ch}$ of Eq.\eqref{eq:rch}, we updated the proton charge distribution to reflect the updated proton charge radius of Ref.~\cite{Pohl:2010}. Particle numbers $N$ and $Z$ are the normalizations of the neutron and proton distributions in Eq.~\eqref{eq:charge}. The ground-state binding energy is calculated from $S^{(p,n)}_{\ell j}(\alpha,\beta;E)$ using the Migdal-Galitski sum rule~\cite{Galitski:1958,Exposed!}. Quasihole energy levels are calculated from a Schr\"odinger-like equation derived from the Dyson equation, see Ref.~\cite{Atkinson:2020} for details.

As discussed previously, the difference between the current fit and the previous one of Ref.~\cite{Mahzoon:2017} involves the inclusion of the CREX-measured $F_W(q^2)$ at $q = 0.8733$ fm$^{-1}$ as a constraint on the DOM self energy. We chose to fit directly to $F_W$ rather than $R_\mathrm{skin}$ to have a closer comparison to what is actually measured. This removes any ambiguity arising from the reduced correlation between $R_\mathrm{skin}$ and $F_W$ in $^{48}$Ca vs. $^{208}$Pb~\cite{Piekarewicz2024}.
Another difference from the earlier fit is the relaxation of the error associated with neutron total cross section data above 100 MeV~\cite{BobCole} that were an important ingredient in generating a thick skin~\cite{Mahzoon:2017}. We found that this adjustment was necessary in order to reproduce a thin skin.

The weak form factor is the Fourier transform of the weak-charge distribution $\rho_{W}(r)$ which, analogously to  $\rho_{ch}(r)$, is calculated by folding the proton and neutron weak-charge distributions with $\rho^{(p,n)}(r)$ from Eq.~\eqref{eq:charge}. Following the prescription of Ref.~\cite{Horowitz:2012}, we also include the spin-orbit contribution which has a non-negligible impact on $F_W$. Notably, we found that explicitly calculating the spin-orbit contributions from all protons and neutrons yielded a different result than calculating only the spin-orbit contribution from the additional $f7/2$ neutrons.  

\begin{figure}[t]
    \includegraphics[width=\linewidth]{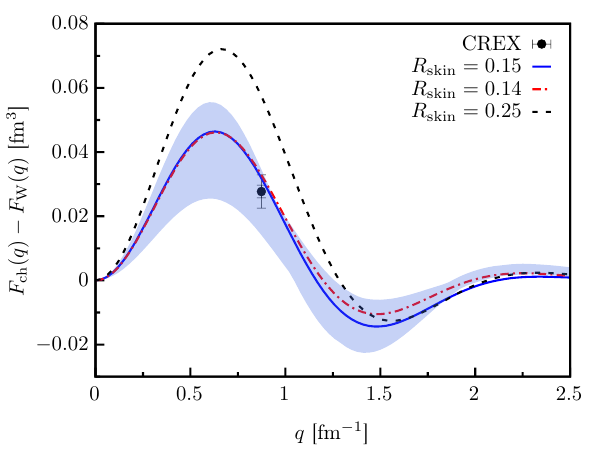}
    \caption{ \label{fig:formf} Difference between the charge and weak form factors in $^{48}$Ca. The data point indicates CREX measurement with both experimental and model uncertainty represented by error bars. The solid blue line is the result of the best DOM fit with $R_\mathrm{skin} = 0.15$ fm while the dash-dotted red line is a comparable DOM fit with $R_\mathrm{skin} =0.14$ fm. The dashed black line is the result of the previous fit from Refs.~\cite{Mahzoon:2017,Atkinson:2019} with $R_\mathrm{skin} = 0.25$ fm. The shaded region is an uncertainty band determined from the experimental errors of the neutron data included in the DOM fit using a bootstrap method.}
\end{figure}

Among the many parametrizations explored during the fitting process, we found several that reproduced similar $F_W$ at the particular value of $q=0.8733$ fm$^{-1}$, but had different $F_W(q)$ shapes for other values of $q$ (see the solid and dot-dashed lines in Fig.~\ref{fig:formf}). 
Included in Fig.~\ref{fig:formf} is also the original DOM prediction of Ref.~\cite{Mahzoon:2017} that yielded a large skin (dashed line).
The two alternative current fits yield skin values of 0.14 and 0.15 fm, where the latter was chosen as the representative fit for this work and subsequent figures where no other fits are indicated. For a brief comparison of these fits, see the supplementary material.
The uncertainty band in Fig.~\ref{fig:formf} was generated by running numerous DOM fits to randomly-scrambled neutron data (including the $F_W$) within 1$\sigma$ of their experimental uncertainties. The standard deviation of this band at $q=0.8733$ fm$^{-1}$ results in an uncertainty of 0.05 fm$^{-1}$ for our predicted weak form factor.
The different weak form factor shapes illustrated in Fig.~\ref{fig:formf} translate to having several different neutron distribution shapes while keeping similar $R_\mathrm{skin}$ values consistent with CREX.  

\begin{figure}[h]
    \includegraphics[width=\linewidth]{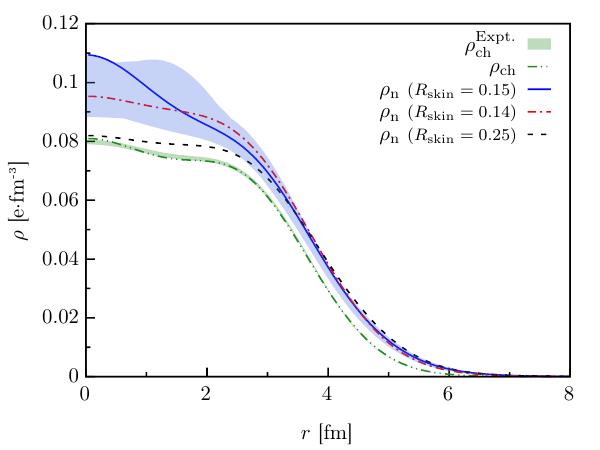}
    \caption{DOM charge and neutron distributions in $^{48}$Ca. The experimental charge distribution is represented by the breen band~\cite{deVries:1987}, with the dash-dot-dot green line representing the charge density from the best-fit DOM with a corresponding $R_\mathrm{skin} = 0.15$ fm. The solid blue line is the best-fit DOM prediction of the neutron distribution of the same fit.  The dash-dotted red line is the neutron distribution of a comparable DOM fit with $R_\mathrm{skin} = 0.14$ fm. The dashed black line is the neutron distribution of our previous DOM fit with $R_\mathrm{skin} = 0.25$ fm~\cite{Mahzoon:2017,Atkinson:2019}. The large shaded region is an uncertainty band for the neutron distribution determined from the experimental errors of the neutron data included in the DOM fit using a bootstrap method.}
    \label{fig:chd-band}
\end{figure}


Additional measurements at different $q$ values can hopefully further constrain the shape of $F_W(q)$ and provide greater insight into the neutron distribution for $^{48}$Ca.
It is clear that measurements of the weak form factor at lower momentum transfer will not affect the skin value (if consistent with CREX) but may reduce the experimental error.
A measurement at higher momentum transfer will of course clarify the properties of the interior weak (and therefore neutron) distribution but will most likely have a substantial percent error, which in turn may not help select the best DOM parametrization. 

It is no surprise that the neutron distributions of the current fits concentrate more strength near $r=0$ fm to accommodate the thin skin of $R_\mathrm{skin} = 0.152$ fm (see Fig.~\ref{fig:chd-band} for a comparison to the previous thick-skin fit).
Due to the Heisenberg uncertainty principle, concentrating neutron presence near the origin leads to increased high-momentum content in the momentum distribution $n(k)$. This high-momentum content is associated with short-range correlation (SRC) pairs. Knockout experiments~\cite {CLAS:2006,Duer:2018} and \textit{ab initio} calculations for asymmetric matter~\cite{Rios:2009,Rios:2014} suggest an increased high-momentum content for the minority species in nuclei. Realistic many-body calculations of low-$A$ nuclei using variational Monte Carlo (VMC) techniques also reveal that the majority of this high-momentum content comes from the tensor force in the nucleon-nucleon interaction~\cite{Wiringa:2014}.

\begin{figure}[h]
    \includegraphics[width=\linewidth]{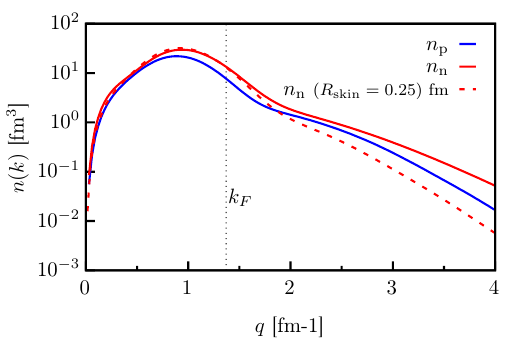}
    \caption{ \label{fig:kcomp} Momentum distribution of neutrons and protons in $^{48}$Ca. The solid blue and red lines are the proton and neutron momentum distributions, respectively, predicted in the current DOM fit with $R_\mathrm{skin} = 0.15$ fm. The dashed red line is the neutron distribution predicted in our earlier DOM fit with $R_\mathrm{skin} =0.25$ fm~\cite{Mahzoon:2017,Atkinson:2019}.}
\end{figure}

\begin{figure}[h]
    \includegraphics{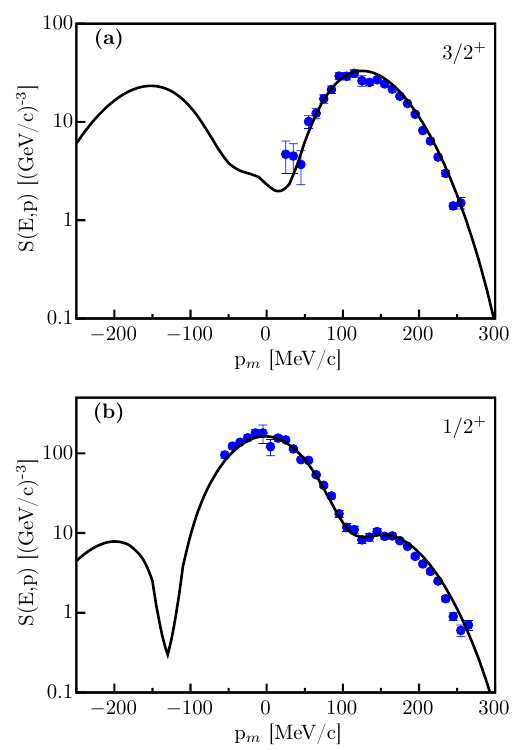}
    \caption{ \label{fig:eep} Comparison of DOM-generated (via DWIA) $^{48}$Ca$(e,e'p)^{47}$K cross sections with experiment at an outgoing proton energy of $100$ MeV. The solid line is the DOM prediction of the current fit while the data points are from Ref.~\cite{Kramer:2001,Kramer90}. (a) Cross section for the removal of a $0$d$3/2$ proton (leaving the resulting $^{39}$K nucleus in a $3/2^+$ state) with a spectroscopic factor of 0.59. (b) Cross section for the removal of a $1$s$1/2$ proton (leaving the resulting $^{39}$K nucleus in a $1/2^+$ state) with a spectroscopic factor of 0.6.}
\end{figure}
The tensor force preferentially acts on neutron-proton ($np$) pairs  with total spin $S=1$. This phenomenon is known as $np$ dominance~\cite{Hen:2017}, and is demonstrated by a factor of 20 difference between the number of observed $np$ SRC pairs and
the number of observed $pp$ and $nn$ SRC pairs in exclusive $(e,e'pp)$ and $(e,e'p)$ cross section measurements of $^{12}$C, $^{27}$Al, $^{56}$Fe, and $^{208}$Pb~\cite{Hen:2017}.
The dominance of $np$ SRC pairs would imply that the number of high-momentum protons observed in a nucleus is dependent on how many neutrons it contains. More specifically, one would expect that the
high-momentum content of protons would increase with neutron excess since there are more neutrons available to make $np$ SRC pairs. The CLAS collaboration confirmed this asymmetry dependence by measuring the
high-momentum content of protons and neutrons from $(e,e'p)$ and $(e,e'n)$ cross section measurements in $^{12}$C, $^{27}$Al, $^{56}$Fe, and $^{208}$Pb~\cite{Duer:2018}.
We note that no such measurements are available for ${}^{48}$Ca.
This nucleus may not conform with expectations as most of the extra neutrons will occupy the valence $f_{7/2}$ orbit.
As it has been demonstrated in the past such an orbit doesn't exhibit high-momentum content itself~\cite{Bobeldijk:1994,Muther:1994}.
The shrinking of the neutron distribution must therefore involve more deeply bound contributions thereby leading to a larger presence of high-momentum components.

Notable is the fact that our previous fit~\cite{Atkinson:2019,Mahzoon:2017} with the thick neutron skin had more high-momentum protons than neutrons. The current fit, due to having more neutrons near the center to accommodate a thin skin, has more high-momentum neutrons than protons instead (see Fig.~\ref{fig:kcomp} for comparison of old and new fits).  
Evidently, the additional experimental point provided by the weak form factor has a critical influence on this change of direction,  solidifying that such data are crucial to provide a complete picture of the neutron properties in $^{48}$Ca. We note that the additional experimental result involves properties of neutrons below the Fermi energy.

Apparently for protons, the available experimental data already provide sufficient constraints to construct such a complete picture. 
The most important data for protons include the charge density and proton reaction cross sections as discussed in Ref.~\cite{Atkinson:2019}. 
Together with level structure near the Fermi energy and differential cross sections in a large energy domain, a sufficiently complete set of constraints is provided to generate predictive power for proton observables that are not part of the fit.
This is illustrated by the success of describing $(e,e'p)$ cross sections as documented in Refs.~\cite{Atkinson:2018,Atkinson:2019}.
The DOM renders the potential for the outgoing proton at the corresponding energy (100 MeV), the overlap function and its spectroscopic factor. Subsequently these ingredients are employed in the distorted wave impulse approximation of the $(e,e'p)$ reaction using the updated DWEEPY code~\cite{Giusti:2011}.
This feature is maintained in the current fit as illustrated in Fig.~\ref{fig:eep}.


The same conclusion cannot be drawn for the neutron data that were originally employed in generating a neutron skin for ${}^{48}$Ca~\cite{Mahzoon:2017}. Even one extra experimental data point that carries information about the neutron distribution, including but not limited to the weak form factor, can alter the picture quite dramatically. It should also be noted that while total neutron cross sections are available, there are no neutron reaction cross sections and only a very small set of elastic scattering data exists~\cite{Mueller:2011}.
Expansion of the experimental data set can therefore contribute significantly to the additional information necessary to pin down the neutron properties in ${}^{48}$Ca.
Our experience with protons suggests that neutron reaction cross sections can provide important constraints~\cite{Atkinson:2019} and it is echoed by the present analysis. 
In fact, the neutron reaction cross section predictions of the $R_\mathrm{skin}=0.15$ fm and $R_\mathrm{skin}=0.14$ fm fits are non-negligibly different, so the addition of these experimental data would help to remove the degeneracy in DOM $F_W$ predictions (see supplementary material).

\begin{figure}[t]
    \includegraphics{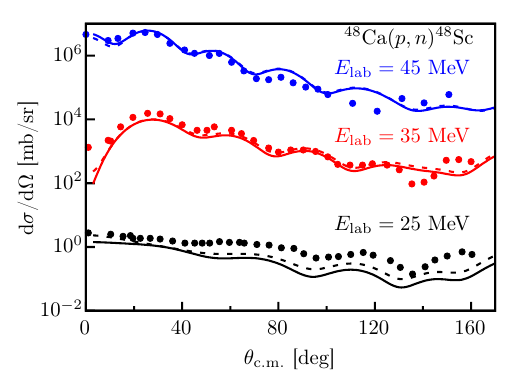}
    \caption{ \label{fig:pn} Comparison of DOM-calculated and experimental charge exchange differential cross sections, $^{48}$Ca$(p,n)^{48}$Sc, at $E_\mathrm{lab} = 25,45,45$ MeV. The points are experimental measurements taken from Ref.~\cite{Doering:1975} while the lines are DOM calculations. Solid  lines are from the current fit with $R_\mathrm{skin} = 0.15$ fm while dashed lines are from the previous fit with $R_\mathrm{skin} = 0.25$ fm ~\cite{Mahzoon:2017,Atkinson:2019}.}
\end{figure}

Another reaction that could help to constrain neutron predictions is the charge exchange reaction, $^{48}$Ca$(p,n)^{48}$Sc, which is largely determined by the isovector component of Lane-like optical potentials~\cite{Pawel17} (such as the DOM). According to Ref.~\cite{Pawel17}, this link with the isovector potential connects charge exchange reactions to neutron skins. While $^{48}$Ca$(p,n)^{48}$Sc data are not included in our fit, we calculated charge-exchange cross sections (in the coupled-channel formalism~\cite{Khoa:2007}) after completion of our fit to determine its sensitivity to our predicted neutron skin. The negligible difference between the charge-exchange predictions of the previous DOM fit with 
$R_\mathrm{skin} = 0.25$ fm and the current fit with $R_\mathrm{skin} = 0.15$ fm (see solid and dashed lines in Fig.~\ref{fig:pn}) indicates that these cross sections are not particularly sensitive to the neutron skin.
Recent work on the $^{48}$Ca$(p,n)^{48}$Sc reaction~\cite{Smith:2024} yields substantial uncertainties which further suggest that no strong conclusions can be extracted concerning the neutron skin from these data.
We note that local optical potentials were employed in the work of Refs.~\cite{Pawel17,Smith:2024} whereas our DOM results are obtained with nonlocal potentials.

\begin{figure}[t]
    \includegraphics[width=\linewidth]{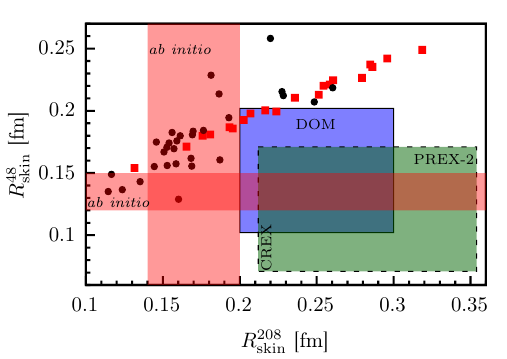}
    \caption{ \label{fig:skins}
   Figure adapted from Ref.~\cite{Horowitz14}. The dashed rectangle represents the CREX and PREX-2 analyses~\cite{crex:2022,prex2:2021}. 
The shaded rectangle labeled DOM represents the DOM results for ${}^{208}$Pb from Ref.~\cite{Atkinson:2020} and the updated $^{48}$Ca from the current fit. Smaller squares and circles refer to relativistic and nonrelativistic mean-field calculations, respectively, cited in Ref.~\cite{Horowitz14}.
The \textit{ab initio} predictions from Ref.~\cite{Hagen:2016} for $^{48}$Ca and Refs.~\cite{Hu:2022,Hu:2024} for $^{208}$Pb are represented by horizontal and vertical bands, respectively, labeled \textit{ab initio}. All uncertainties are reported at the $1\sigma$ level.
   }
\end{figure}

The current results for the DOM neutron skins in ${}^{48}$Ca and ${}^{208}$Pb are summarized in Fig.~\ref{fig:skins}.
As demonstrated above, our constrained self-energies for $^{48}$Ca utilize both scattering and bound-state data for a robust picture of nuclei. Such a fit for ${}^{48}$Ca now includes the CREX $F_W$ data point resulting in a thin neutron skin, $R^{\mathrm{DOM}48}_\mathrm{skin}=0.15 \pm 0.05$ fm, while the corresponding result for ${}^{208}$Pb, remains at $R^{\mathrm{DOM}208}_\mathrm{skin}=0.25 \pm 0.05$ fm using the uncertainty quantification clarified in Refs.~\cite{Mahzoon:2017,Atkinson:2020}. 
These results are represented by the shaded box labeled DOM in Fig.~\ref{fig:skins}. 
The figure is adapted from Ref.~\cite{Horowitz14} and includes the coupled-cluster result from Ref.~\cite{Hagen:2016} as a horizontal band.
The vertical band represents the \textit{ab initio} work reported in Refs.~\cite{Hu:2022,Hu:2024}.
Relativistic and nonrelativistic mean-field calculations cited in Ref.~\cite{Horowitz14} are represented by squares and circles, respectively.
The dashed rectangle is centered on the CREX and PREX-2 results. As made evident by the overlapping of the DOM and CREX-PREX boxes in Fig.~\ref{fig:skins}, the DOM can now simultaneously describe both the thin CREX and thick PREX-2 neutron skins.



In conclusion, we have updated our DOM fit of $^{48}$Ca to reproduce the CREX measurement of the $^{48}$Ca weak form factor and its correspondingly neutron skin. A natural consequence of predicting a thin neutron skin is a neutron distribution with concentrated strength in the interior of the nucleus. This translates to more high-momentum neutrons reversing the hierarchy from the previous thick-skin fit, which could point to some tension with the $np$ dominance observed in CLAS measurements of asymmetric nuclei~\cite{CLAS:2006,Duer:2018}. 

We have demonstrated that the availability of more neutron scattering and bound-state data would allow for a more precise description of $^{48}$Ca and would also elucidate the DOM predictive power for neutron observables, leveling it to its current capability with respect to protons. 
Furthermore, constraining $F_W$ to only a single $q$ value allows ambiguity in the shape of weak form factor $F_W$ and hence the neutron distribution. Ideally, future measurements (such as the Mainz Radius Experiment (MREX)) will be performed at different $q$ values to pinpoint a more detailed shape of $F_W$ and hence the neutron distribution.


This work was performed under the auspices of the U.S. Department of Energy by Lawrence Livermore National Laboratory under Contract DE-AC52-07NA27344 and was supported by the LLNL-LDRD Program under Project No. 24-LW-062.
This work was also supported by the U.S. National Science Foundation under grants PHY-1912643 and PHY-2207756.

\bibliographystyle{apsrev4-1}
\bibliography{newskin}

\end{document}